\documentclass[conference]{IEEEtran}
\IEEEoverridecommandlockouts
\usepackage{cite}
\usepackage{amsmath,amssymb,amsfonts}
\usepackage{graphicx}
\usepackage{textcomp}
\usepackage{xcolor}
\def\BibTeX{{\rm B\kern-.05em{\sc i\kern-.025em b}\kern-.08em
    T\kern-.1667em\lower.7ex\hbox{E}\kern-.125emX}}

\usepackage{subcaption}
\usepackage[utf8]{inputenc}
\usepackage[T1]{fontenc}
\usepackage{booktabs}
\usepackage{tabularx}
\usepackage{url}
\DeclareMathOperator*{\argmin}{arg\,min}

\usepackage{tikz}

\usepackage{algorithm}
\usepackage{algpseudocode}

\title{Memory-Efficient RkNN Retrieval by Nonlinear k-Distance Approximation}

\author{\IEEEauthorblockN{Sandra Obermeier}
\IEEEauthorblockA{\textit{LMU Munich}\\
Munich, Germany \\
obermeier@dbs.ifi.lmu.de}
\and
\IEEEauthorblockN{Max Berrendorf}
\IEEEauthorblockA{\textit{LMU Munich}\\
Munich, Germany \\
berrendor@dbs.ifi.lmu.de}
\and
\IEEEauthorblockN{Peer Kröger}
\IEEEauthorblockA{\textit{LMU Munich}\\
Munich, Germany \\
kroeger@dbs.ifi.lmu.de}
}

\begin{document}

\maketitle

\begin{abstract}
The reverse k-nearest neighbor (RkNN) query is an established query type with various applications reaching from identifying highly influential objects over incrementally updating kNN graphs to optimizing sensor communication and outlier detection.
State-of-the-art solutions exploit that the k-distances in real-world datasets often follow the power-law distribution, and bound them with linear lines in log-log space.
In this work, we investigate this assumption and uncover that it is violated in regions of changing density, which we show are typical for real-life datasets.
Towards a generic solution, we pose the estimation of k-distances as a regression problem.
Thereby, we enable harnessing the power of the abundance of available Machine Learning models and profiting from their advancement.
We propose a flexible approach which allows steering the performance-memory consumption trade-off, and in particular to find good solutions with a fixed memory budget crucial in the context of edge computing.
Moreover, we show how to obtain and improve guaranteed bounds essential to exact query processing.
In experiments on real-world datasets, we demonstrate how this framework can significantly reduce the index memory consumption, and strongly reduce the candidate set size.
We publish our code at \url{https://github.com/sobermeier/nonlinear-kdist}.
 \end{abstract}


\section{Introduction}
\begin{figure}
    \centering
    \includegraphics[width=.8\linewidth]{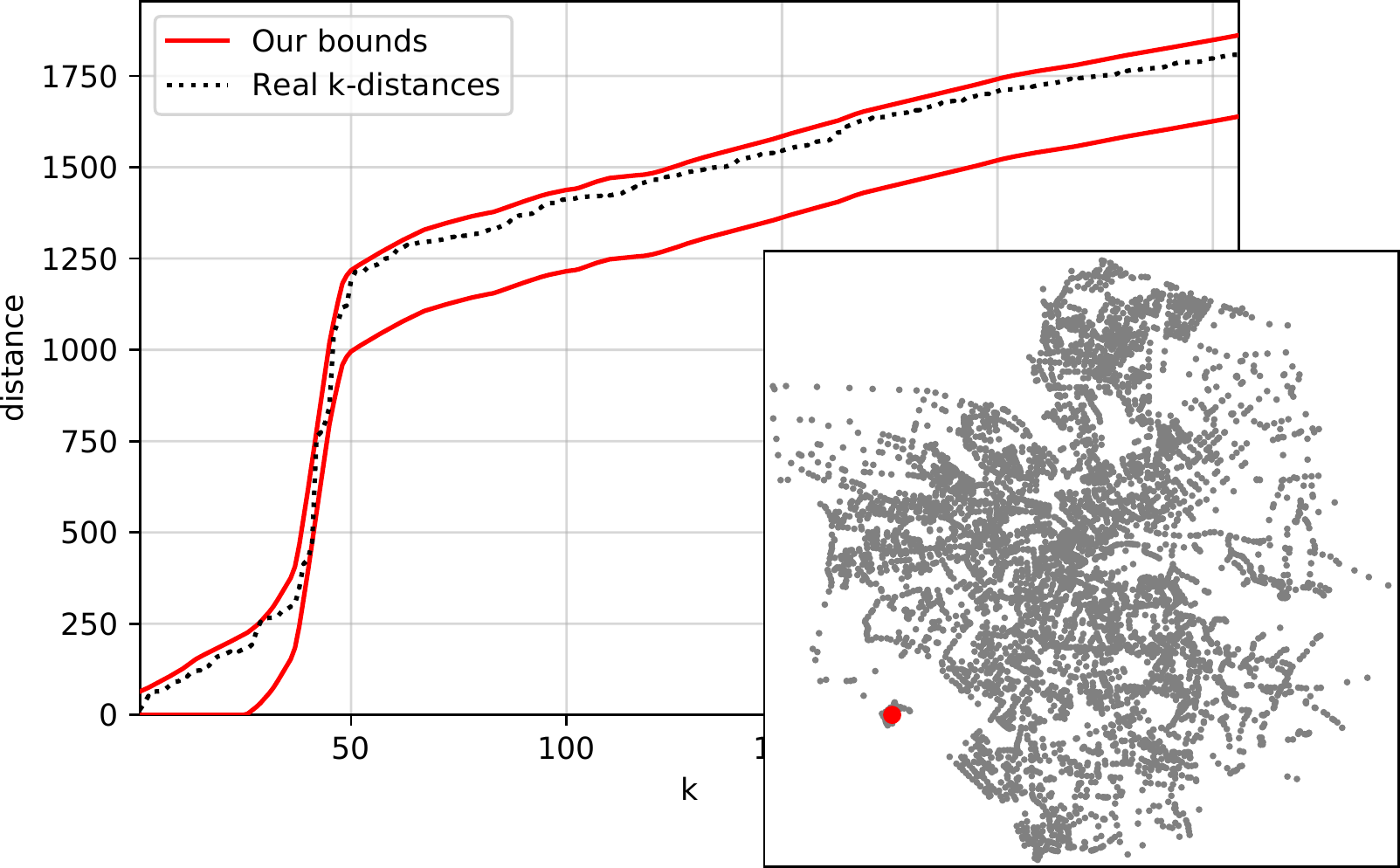}
    \caption{
$k$-distance (y-axis) for a selected point (red) in a toy dataset (small inset) and different values of $k$ (x-axis).
The $k$-distances exhibit non-linear behavior with steep increase at the borders of dense regions.
    }
    \label{fig:teaser}
\end{figure}

Besides range and k-nearest neighbor (kNN) queries, another frequently used query type is reverse k-nearest neighbor (RkNN) queries given as $RkNN(q) = \{o \in D \mid q \in kNN(o)\}$.
Thus, it returns those objects in the database $D$, for which the query is one of the $k$-nearest neighbors.
This task is different to the kNN query, since the $k$-nearest neighbor relation is \emph{not} symmetric.
Consider a toy dataset of numbers $\{1,2,4\}$ with the absolute difference as distance.
Then, 2 is the nearest neighbor of 4, but not vice-versa.
As na\"ively $\mathcal{O}(n)$ kNN queries are necessary to determine the result, numerous approaches for efficient query processing and effective indexing have been proposed~\cite{Achtert:2006:ERK:1142473.1142531,wu2008finch,cheema2011influence,casanova2017dimensional,yang2017reverse}.
The query result itself describes the influence set of an object and is hence of interest itself.
However, it has further applications in fields reaching from outlier detection~\cite{lijun2010data} over clustering~\cite{hu2008hypergraph,8942511} to incremental updates for kNN query processing in sensor networks~\cite{DBLP:journals/percom/YangC17}. 
Existing real-world applications are often limited by the given storage capacity, especially when considering modern, embedded systems. 
State-of-the-art approaches are often situated within the filter-refinement framework.
Here, a fast \emph{filter} method is applied which decides for a large proportion of the data whether they are certainly part of the result (\emph{safe inclusion}) or certainly not part of the result (\emph{safe exclusion}).
Afterward, only for the few remaining data points for which no decision could be taken, the expensive forward $k$NN queries are performed.
Thus, research focuses mainly on deriving new efficient and effective filter methods.
In this work, we investigate how to derive filters using Machine Learning models that approximate the k-distances of each point.
To this end, we pose the regression problem to predict the $k$-distance $nndist(p, k)$ of a data point given its representation $p$.
Moreover, we show how to adapt the principles of learned index structures \cite{Kraska:2018:CLI:3183713.3196909} to derive guaranteed lower and upper bounds.
In our experiments, we demonstrate how the approach can be applied to several real-world datasets reducing the number of candidates as well as the index size.

In summary, our contributions are:
\begin{itemize}
    \item We propose a framework to learn the distribution of $k$-distances as a regression task given an object's representation. By employing powerful Machine Learning models, this framework can model varying densities.
    \item We derive and enhance guaranteed bounds essential for the application to exact query processing within filter-refinement.
    \item We conduct a thorough empirical study to evaluate our proposed approach. The results show both a reduction in index size and in candidate set size.
\end{itemize}

The remainder of this paper is structured as follows:
In Section~\ref{sec:related_work}, we review existing work both in the field of RkNN query processing as well as learned index structures and position our work.
In Section~\ref{sec:contribution}, we formalize the regression problem, and elaborate on how to obtain and improve guaranteed lower and upper bounds for exact query processing,
In Section~\ref{sec:experiments}, we present an extensive experimental evaluation analyzing the trade-off between model size and candidate set sizes, as well as performing an ablation study to investigate the effect of individual components. 
In Section~\ref{sec:conclusion}, we conclude this paper and outline possible directions for future research. 
\section{Related Work}
\label{sec:related_work}

\subsection{RkNN Query Processing}
\label{sbusec:rknn-query-processing}
In general, there are \emph{static} and \emph{dynamic} RkNN query processing approaches.
While the first assume a static set of objects in a database, the latter can cope with incremental changes such as insertion and deletion of data points.
Dynamic approaches can also be applied to a static database, but they might lose some efficiency as they cannot exploit the fact that all objects are already known and there will be no changes.
A common approach to applying static methods in a dynamic setting is to keep an additional database that contains only the changes.
Query Processing is then performed on the indexed database with index support and in a sequential search in the database of changes.
From time to time the static index is rebuilt to include the changes that happened so far.
This is often a viable alternative for slowly changing databases.
Thus, in our work, we focus only on a $\emph{static}$ environment.

Besides, we can distinguish two variants of use-cases for RkNN query processing: the $\emph{mono-chromatic}$ and the $\emph{bi-chromatic}$ case \cite{Korn:2000:ISB:342009.335415}.
In the $\emph{bi-chromatic}$ case, the query objects, and data objects belong to two different categories, such as service providers and customers.
However, in our experiments, we focus on the  $\emph{monochromatic}$ version where the query objects belong to the same category as the data objects.

\subsubsection{R-Tree Extensions}
The first approaches on RkNN query processing were based on an extension of the R-Tree where objects are saved in an R-Tree in a specific form. 

The \emph{RNN-Tree}\cite{Korn:2000:ISB:342009.335415} is only applicable for RkNN queries with $k = 1$.
Spheres in the form of $(p, nndist(p,1))$ are stored in an R-Tree for each data point $p \in D$, with center $p$ and radius of $nndist(p,1)$, i.e. the $k$-distance for $k=1$.
A query point $q$ only influences data points that contain $q$ because only these points have $q$ as their nearest neighbor.
So for a query $q$ the centers $p$ of all data objects are returned where the circle $(p, nndist(p,1))$ contains $q$.
Furthermore, the RNN Tree is not very efficient as there are many leaf accesses required for locating the RNNs. 

The \emph{RdNN-Tree}\cite{914862} stores extra information about the nearest neighbors of the points directly in the R-tree and can handle both NN queries and RNN queries. 
Still only $k=1$ is supported.
Like the RNN-Tree, the leaf nodes store entries of the form $(p,nndist(p,1))$. 
The inner nodes, however, store an array of branches in the form of $(ptr, Rect, max\_nndist)$.
$ptr$ points to another node in the tree, either a leaf or an inner node.
$Rect$ is the minimum bounding rectangle of all elements stored in the sub-tree.
$max\_nndist$ refers to the maximum k-distance of all points $p_s$ contained in the sub-tree of this node.
An RNN search for a query point $q$ is done as follows:
For a leaf node, each point $p$ in the node is examined. 
If $dist(p,q) \leq nndist(p,1)$ then $p$ belongs to the R1NN of $q$.
For an intermediate node, $q$ is compared to the branches. 
If $dist(q, Rect) > max\_nndist$ then the branch can be pruned because all data points within the sub-tree rooted at the investigated branch have a smaller distance to their nearest neighbor. 

\subsubsection{Filter-Refinement Approaches}
Later, more efficient approaches based on a \emph{filter refinement architecture} were proposed.
RkNN query processing is divided into two phases, the \emph{filter} and the \emph{refinement} phase.
In the filter step, we run a fast procedure to exclude as many objects as possible.
For exact query processing we must ensure that no object which is part of the result is discarded.
The remaining objects, called \emph{candidates}, go through a refinement phase. 
In the refinement phase, for each candidate an expensive kNN query is performed to decide whether it is part of the result.

\emph{TPL}\cite{Tao:2004:RKS:1316689.1316754} is an algorithm which is based on the filter refinement architecture.
First, the entries of an R-Tree are traversed into a heap ordering them by ascending distance to the query point $q$ because RNNs are likely to be near $q$. 
In the filter step, the algorithm goes through the heap and uses a concept of half-planes to prune data objects that cannot be candidates. 
Considering the perpendicular bisector between a query point $q$ and an arbitrary data object $p$ that divides the data space into two half-planes $PL_q(p,q)$ containing $q$ and $PL_p(p,q)$ containing $p$. 
Every object, a single data object or a minimum bounding rectangle containing many data objects, that is in the $PL_p(p,q)$ cannot be an RNN of q because it is closer to $p$. 
This is the basic idea behind the pruning of the TPL algorithm. 
Even though originally it was proposed for R1NN queries the paper of \cite{Tao:2004:RKS:1316689.1316754} also presents a way to extend TPL for RkNN queries. 
The data objects determined as non-candidates are not discarded immediately. 
They are added to a set called $S_{rfn}$. 
Data objects that cannot be pruned are added to the candidates set $S_{cnd}$.
In the refinement step entries of this $S_{rfn}$ are used to identify false hits in the $S_{cnd}$. 

The \emph{MRkNNCoP Tree}\cite{Achtert:2006:ERK:1142473.1142531}
is another example within the filter refinement framework. 
Lower and upper bounds of the k-distance are used as a filter. 
A novel generic index for the RkNN search is proposed based on the ideas of RdNN Tree. 
Instead of an R-Tree, an M-Tree\cite{article} is used to generalize from Euclidean vector data to metric objects. 
The MRkNNCoP Tree introduces another improvement to the RdNN Tree by generalizing $k$.
The MRkNNCoP approach exploits the observation that the distribution of distances for a specific point in a natural dataset often follows the power-law. 
Thus, for each point, the lower and the upper bound of the $k$-distance are approximated by a linear line in log-log space, thus requiring only two parameters per bound, i.e. four parameters per data point.
An extended M-Tree is used for aggregating the maximum of all upper bounds and the minimum of all lower bounds for each node and all data objects contained in that node. 

\subsection{Learned Index Structure (LIS)}
\label{sec:lis}
The basic idea of \emph{Learned Index Structures}\cite{Kraska:2018:CLI:3183713.3196909} is that indexes can be seen as models.
More concrete, the authors show how a B-Tree, a hash map, or a Bloom filter can be modeled as regression, or classification model, either mapping a key to a position or predict its existence.
For example, a B-Tree can be considered as a model that maps a key to the position of a specific value in a key-sorted list.
Furthermore, in this setup only the first key of a page is indexed for efficiency reasons. So, it is guaranteed that the key of the record at the position given by the B-Tree is the first key equal ($pos + 0$) or higher ($pos + pagesize$) than the look-up key.
Hence, the B-Tree is a model with a guaranteed minimal and maximal error. 
Within a static setting, the minimal error and maximal error of a prediction can also be guaranteed and thus, also the range of a predicted position.
The minimal position is $pos-\Delta_\downarrow$ where $\Delta_\downarrow$ is the minimal difference between the prediction and the true position and the maximal position is determined by $pos+\Delta_\uparrow$ where $\Delta_\uparrow$ is the maximum difference between the prediction and the true position.
The main advantage of learned indexes is that the cost of a lookup operation is reduced from $\mathcal{O}(\log n)$ to $\mathcal{O}(1)$, and the storage costs are reduced from $\mathcal{O}(n)$ to $\mathcal{O}(1)$.

\cite{yu2015neural,Gripon2018} propose to accelerate (approximated) nearest neighbor search by using associative memories, and are hence related to learned index structures.
During indexing time, the data is partitioned into equi-sized bins.
During query time, only those bins are refined having the largest overlap with the query object.
In \cite{Oosterhuis2018}, learned indexed for conjunctive boolean queries are considered, building upon the work of \cite{Kraska:2018:CLI:3183713.3196909}.
To this end, they propose multiple approaches each having a different trade-off between storage requirements and computational effort.
The approaches are applied to term-queries for a document database.
Xiang et al. \cite{Xiang2019} propose the Pavo index, which is a learned inverted index.
The underlying hash function is replaced by a hierarchy of recurrent neural network models.
Dong et al. \cite{Dong2020Learning} propose Neural LSH for fast approximate nearest neighbor search.
From the data points, a kNN graph is built, and a partition thereof is learned.
There the first level comprises a NN predicting the partition index.
This partition is then further split using k-means.

However, the existing approaches for (approximate) kNN query processing with learned indices \cite{yu2015neural,Gripon2018,Dong2020Learning}, are not directly efficiently applicable: 
Instead of determining the exact result of the kNN query, it suffices to decide whether $q$ belongs to it.

\subsection{LIS for RkNN}
\cite{DBLP:conf/sisap/BerrendorfBK19} propose an approach to reduce the memory requirements of the MRkNNCoP tree.
The parameters of the log-log linear bounds of the MRkNNCoP tree are approximated by various Machine Learning regression models.
Guaranteed bounds are subsequently derived by using the minimum and maximum training error.
Despite appealing compression ratios, due to the "double approximation" of first approximating the coefficients of a model, which then, in turn, approximates the k-distance, this approach inherits the limitation of MRkNNCoP tree to be only able to model linear bounds in log-log space and sacrifices some of its performance for smaller index sizes.
We address this issue of having two approximation steps each with its loss of precision by combining them into a single, direct approximation of the k-distances.
 
\section{Proposed Method}
\label{sec:contribution}
Within this paper we use the following notation: 
Let $(U, dist)$ be a distance space, i.e. $U$ is an arbitrary set and $dist: U \times U \to \mathbb{R}$ is a distance, i.e. it fulfills the following axioms for all $x,y \in U$ \cite{DBLP:books/daglib/0015656}:
\begin{enumerate}
    \item Non-Negativity: $dist(x,y) \geq 0$
    \item Symmetry: $dist(x,y) = dist(y,x)$
    \item $dist(x,x) = 0$
\end{enumerate}
Furthermore, let $D \subseteq U, |U| < \infty$ be a database, $k \in \mathbb{N}$ and $q \in U$ be a query point.
By 
\begin{equation}
    nndist(x, k) := \argmin \limits_{\substack{D' \subset D\\|D'|=k}} \max \limits_{x' \in D} dist(x, x')
\end{equation}
we denote the $k$-distance of an object $x$, i.e. the distance to its $k$-nearest neighbor.

\begin{algorithm}
\begin{algorithmic}
\Function{RNN}{q,k}
\State $result, candidates \gets \emptyset, \emptyset$
\ForAll{$o \in D$} \Comment{Filter Step}
\State $d[o] \gets distance(q, o)$
\State $upper, lower \gets \text{getBounds}(o, k)$
\If {$d < lower$}
    \State $result \gets results \cup \{o\}$
\ElsIf {$d < upper$}
    \State $candidates \gets candidates \cup \{o\}$
\Else
    \State Reject $o$
\EndIf
\ForAll{$o \in candidates$} \Comment{Refinement Step}
\State $kd \gets nndist(o, k)$
\If{$d[o] < kd$}
    \State $result \gets result \cup \{o\}$
\Else
    \State Reject $o$
\EndIf
\EndFor
\EndFor
\State \Return $result$
\EndFunction
\end{algorithmic}
\caption{RkNN query processing within the filter-refinement framework.}
\label{alg:framework}
\end{algorithm}
We briefly revisit the filter-refinement framework for processing RkNN queries.
The general pipeline is given in Algorithm~\ref{alg:framework}.
Given a query point $q$ and the parameter of $k$, the algorithm proceeds by first screening the whole database, and calculating the distance between the query point, and the object $o$.
Besides, the filter bounds are calculated, which give lower and upper bounds of the true $k$-distance.
If the distance is smaller than the lower bound, it is also smaller than the true $k$-distance, and hence can immediately be included in the result without further verification.
If the distance is larger than the upper bound, it will also be larger than the true $k$-distance, and hence can directly be discarded.
For those in between, we cannot decide without further calculation and hence store them into a set of \emph{candidates}.
After the filter step is finished, we have a partial result, and well as a set of candidates.
For this set of candidates we have to compute the real $k$-distances, which each requires an expensive forward $k$NN query.

\begin{figure}
  \begin{center}
   \includegraphics[width=.7\linewidth, height=.6cm, keepaspectratio]{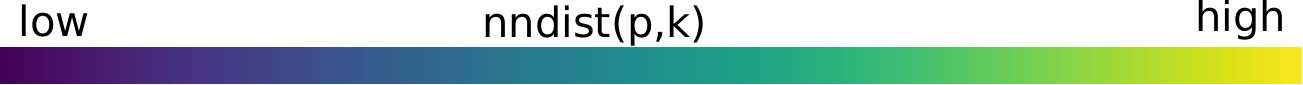}\\
   \includegraphics[width=0.4\linewidth]{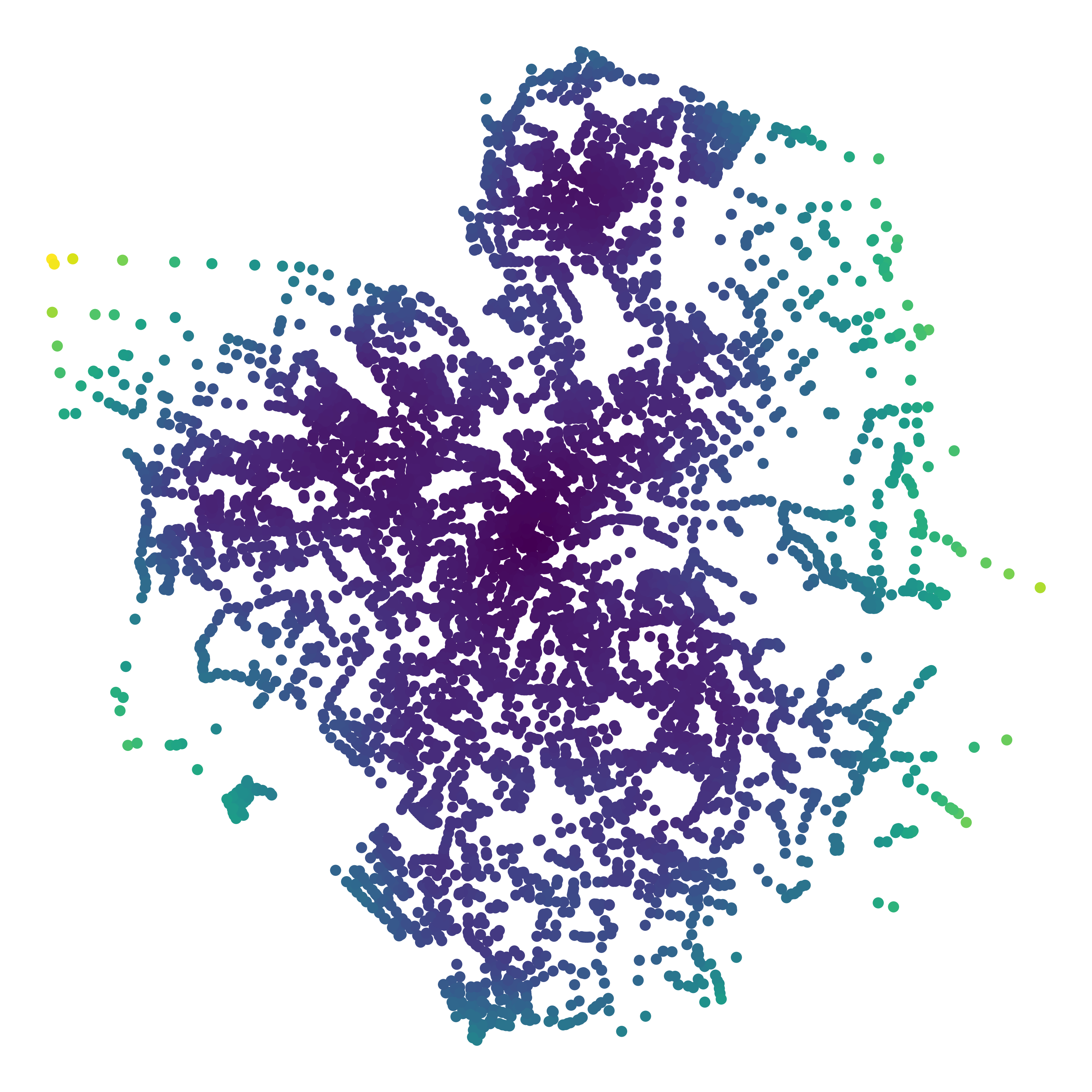}
   \includegraphics[width=0.4\linewidth]{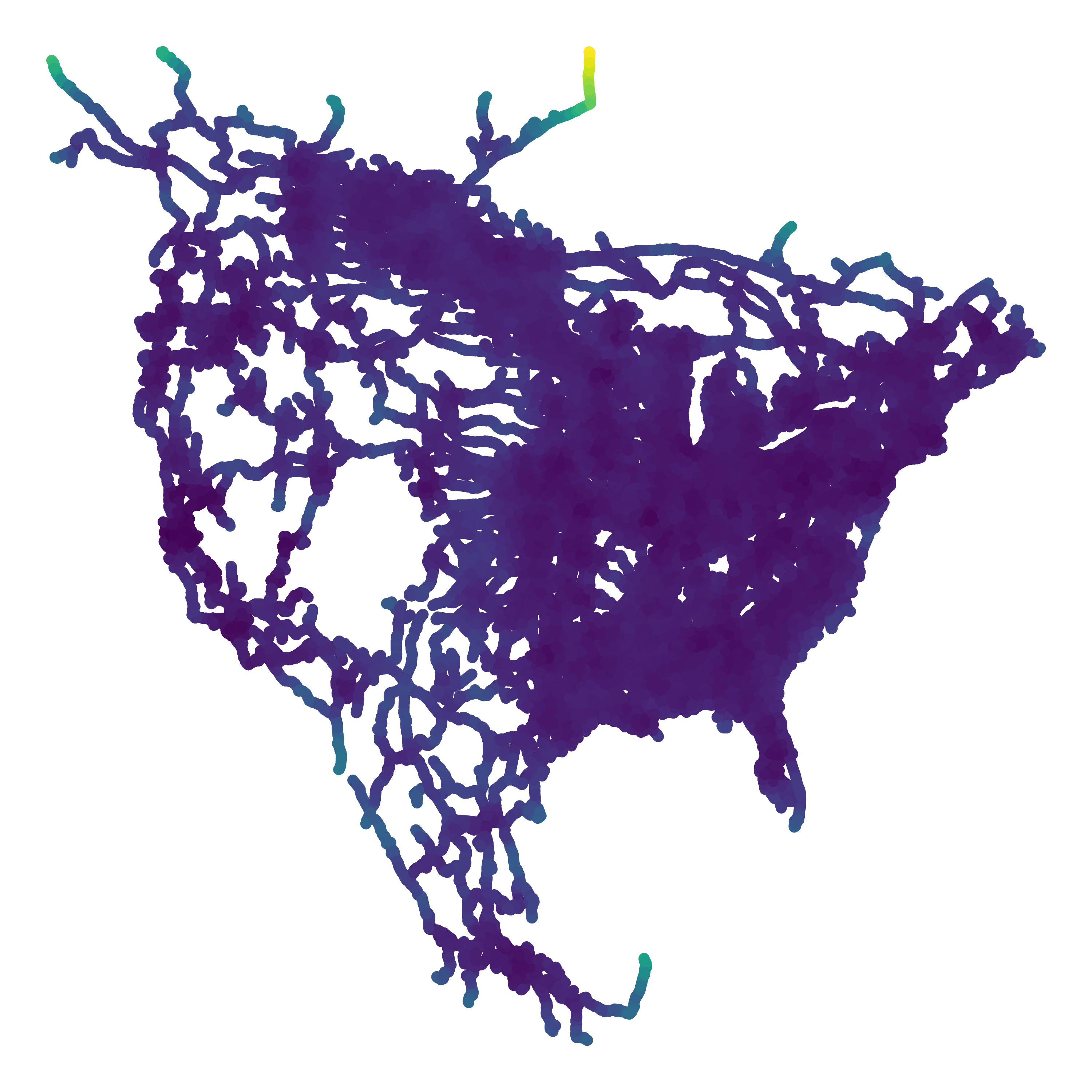}
  \end{center}
  \caption{k-distances mapped on road networks Oldenburg (left) and North America (right) for $k=256$. In dense regions the k-distances are smaller while they are larger for sparse regions.
  \label{fig:skd-ol-na-256}
  }
\end{figure}
Towards these bounds, we pose the regression prediction of the (continuously valued) $k$-distance of a data point given its representation, i.e. we seek regression model $M$ with parameters  $\theta$ such that $M(x, k; \theta) \approx nndist(x, k)$.
Our motivation to use the object's representation as input to our model is that the distances are also calculated based upon these representations.
To undermine this assumption consider the $k$-distances within a real world dataset.
Figure~\ref{fig:skd-ol-na-256} shows the k-distances for $k=256$ mapped onto the road networks of Oldenburg and North America. 
It can be observed that data points that are close have similar k-distances and that the k-distances are smaller in dense regions.
We hypothesize that this kind of correlation between coordinates of the data point, and its $k$-distances can be captured by a Machine Learning model.

Depending on the type of model we can encode different inductive biases.
For instance, tree-based models such as regression trees learn a hierarchical split of the data space and predict separate values for each of these.
Thus, they resemble classical tree-based index structures, such as R-Trees.
We refer to \cite{Kraska:2018:CLI:3183713.3196909} for further explanation on how e.g. B-Trees can be seen as regression models.
Neural networks on the other hand are universal function approximators, and depending on the activation function, learn functions that resemble piece-wise linear functions \cite{DBLP:conf/nips/LuPWH017}.

For a discussion about how to transform a trained Machine Learning model into an optimized C code, and thereby achieve efficiency, we refer to \cite{Kraska:2018:CLI:3183713.3196909}.
Moreover, our models do easily fit into RAM and are a direct mapping of the object's representation to the $k$-distance, which addresses the indexability.
Thus, the subsequent section focuses on how to obtain \emph{selective} and \emph{complete} bounds.

\subsection{Guaranteeing Completeness}
\label{sec:bounds-calc}
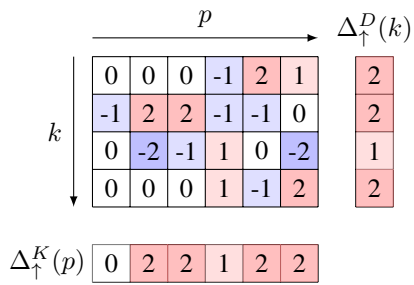
\begin{figure}
    \centering
    \begin{tikzpicture}[scale=.5]
    \colorlet{0}{blue!50}
    \colorlet{1}{blue!25}
    \colorlet{2}{white}
    \colorlet{3}{red!25}
    \colorlet{4}{red!50}
    
    \foreach \x/\y/\c/\v in {0/0/2/0,0/1/2/0,0/2/1/-1,0/3/2/0,1/0/2/0,1/1/0/-2,1/2/4/2,1/3/2/0,2/0/2/0,2/1/1/-1,2/2/4/2,2/3/2/0,3/0/3/1,3/1/3/1,3/2/1/-1,3/3/1/-1,4/0/1/-1,4/1/2/0,4/2/1/-1,4/3/4/2,5/0/4/2,5/1/0/-2,5/2/2/0,5/3/3/1}{
        \fill[color=\c, opacity=.5] (\x,\y) rectangle +(1, 1);
        \node[xshift=.25cm, yshift=.25cm] at (\x,\y) {\v};
    }
    \draw (0, 0) grid (6, 4);
    
    \node at (3, 5) {$p$};
    \node at (-1, 2) {$k$};
    
    \draw[-latex] (0, 4.5) -- +(6, 0);
    \draw[-latex] (-.5, 4) -- +(0, -4);
    
    \draw (0, -1) grid +(6, -1);
    \foreach \x/\c/\v in {0/2/0,1/4/2,2/4/2,3/3/1,4/4/2,5/4/2}{
        \fill[color=\c, opacity=.5] (\x,-2) rectangle +(1, 1);
        \node[xshift=.25cm, yshift=.25cm] at (\x,-2) {\v};
    }
    \draw (-1.2, -1.5) node {$\Delta_\uparrow^K(p)$};
    
    \draw (7, 0) grid +(1, 4);
    \foreach \y/\c/\v in {0/4/2,1/3/1,2/4/2,3/4/2}{
        \fill[color=\c, opacity=.5] (7,\y) rectangle +(1, 1);
        \node[xshift=.25cm, yshift=.25cm] at (7,\y) {\v};
    }
    \draw (7.5, 4.7) node {$\Delta_\uparrow^D(k)$};
    \end{tikzpicture}
    \caption{
    Toy example to illustrate different aggregation modes.
    The matrix shows the residuals, i.e. difference between the true $k$-distance and the prediction, for $k=1,\ldots,4$ and 6 different points.
    By max-aggregation over one of the axes we obtain the maximum residuals which can be used for upper-bounding the real $k$-distance with reduced storage requirements.
    }
    \label{fig:my_label}
\end{figure}
Let $M$ be a model predicting the k-distances. We calculate the residual $
    \Delta(p,k) = nndist(p,k) - M(p,k)
$
If we can bound the residual from above and below we can derive guaranteed bounds for the $k$-distance as well, i.e. if
$
\Delta_\downarrow(p, k) \leq \Delta(p,k) \leq \Delta_\uparrow(p, k)
$
we have
$
lb(p, k) := M(p,k) + \Delta_\downarrow(p, k)
\leq nndist(p,k) 
\leq M(p,k) + \Delta_\uparrow(p, k) := ub(p,k)
$
These bounds on the residual are essential as they guarantee the completeness of the filter step, i.e. that we do not discard any object which should be part of the result.
If we use the lower bound to include objects into the result without explicit $k$-distance computation, the guaranteed lower bound is also crucial for the correctness of the algorithm.

While numerous ways of obtaining such bounds are possible, in this work, we investigate two simple choices:
We calculate the residuals for all data objects $p$ and all values of $k=1,\ldots,k_{max}$, and fixing either $k$ or $p$ and aggregating over the other parameter.
Thereby, we can achieve either $\mathcal{O}(k_{max})$ or $\mathcal{O}(n)$, which may differ in their selectivity.

\subsubsection{Aggregating over $p$}
In this case, for each fixed $k$, we aggregate the residual over all $p' \in D$, and hence obtain bounds with additional storage overhead of $(2 \cdot k) \in \mathcal{O}(k_{max})$.
\begin{eqnarray}
\Delta_\downarrow^D(p, k) &=& \Delta_\downarrow^D(k) = \min_{p' \in D} \Delta(p', k)\\
\Delta_\uparrow^D(p, k) &=& \Delta_\uparrow^D(k) = \max_{p' \in D} \Delta(p', k)
\end{eqnarray}
Thereby, for a fixed value of $k$, the width of the bounds is equal for all points.

\subsubsection{Aggregating over $k$}
Here, for each data point individually, we aggregate the residuals over all values of $k$.
Thus, the bounds require an additional storage space of $(2 \cdot n) \in \mathcal{O}(n)$.
\begin{eqnarray}
\Delta_\downarrow^K(p, k) &=& \Delta_\downarrow^K(p) = \min_{1 \leq k' \leq k_{max}} \Delta(p, k')\\
\Delta_\uparrow^K(p, k) &=& \Delta_\uparrow^K(p) = \max_{1 \leq k' \leq k_{max}} \Delta(p, k')
\end{eqnarray}
Thereby, for a fixed point $p$, the width of the bounds is equal over all values of $k$.

\subsubsection{Combination of Aggregations}
We can obtain guaranteed bounds with memory requirements of $\mathcal{O}(k_{max})$ or $\mathcal{O}(n)$ by aggregating residuals.
Since both a guaranteed bounds, we may do both and combine them to new improved bounds at cost of  $\mathcal{O}(n +k_{max})$.
\begin{eqnarray}
\Delta_\downarrow^{KD}(p, k) &=& \max \{\Delta_\downarrow^K(p), \Delta_\downarrow^D(k)\}\\
\Delta_\uparrow^{KD}(p, k) &=& \min \{\Delta_\uparrow^K(p), \Delta_\uparrow^D(k)\}
\end{eqnarray}

\subsection{Increasing Selectivity}
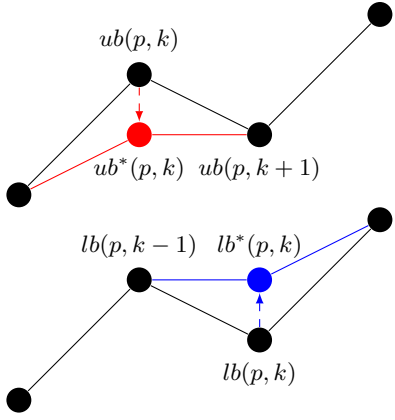
\begin{figure}
\centering
\begin{subfigure}{\linewidth}
  \centering
  \begin{tikzpicture}[
  every node/.style={circle,fill},
  every label/.append style={rectangle, font=\small},
  xscale=2,
  scale=.8,
  ]
  \draw (0, 0) node (a) {};
  \draw (1, 2) node[label={$ub(p,k)$}] (b) {};
  \draw (1, 1) node[color=red,label=below:{$ub^*(p,k)$}] (b') {};
  \draw (2, 1) node[label=below:{$ub(p,k+1)$}] (c) {};
  \draw (3, 3) node (d) {};
  \draw (a) edge (b);
  \draw (b) edge (c);
  \draw (c) edge (d);
  \draw (a) edge[red] (b');
  \draw (b') edge[red] (c);
  \draw (b) edge[red, -latex, dashed] (b');
  \end{tikzpicture}
  \label{fig:mono-ub}
\end{subfigure}
\begin{subfigure}{\linewidth}
  \centering
  \begin{tikzpicture}[
  every node/.style={circle,fill},
  every label/.append style={rectangle, font=\small},
  xscale=2,
  scale=.8,
  ]
  \draw (0, 0) node (a) {};
  \draw (1, 2) node[label={$lb(p,k-1)$}] (b) {};
  \draw (2, 1) node[label=below:{$lb(p,k)$}] (c) {};
  \draw (2, 2) node[color=blue,label=above:{$lb^*(p,k)$}] (c') {};
  \draw (3, 3) node (d) {};
  \draw (a) edge (b);
  \draw (b) edge (c);
  \draw (c) edge (d);
  \draw (b) edge[blue] (c');
  \draw (c') edge[blue] (d);
  \draw (c) edge[blue, -latex, dashed] (c');
  \end{tikzpicture}
  \label{fig:mono-lb}
\end{subfigure}
\caption{Visualization of restoring monotonicity for upper and lower bound exploiting monotonicity.}
\label{fig:mono-ub-lb}
\end{figure}
After obtaining the bounds, we can exploit two properties of the k-distances, namely that k-distances are non-negative and monotonous, in a parameter-free postprocessing step to enhance predictions and bounds.

\subsubsection{Non-Negativity}
As all distances, the k-distances are non-negative. Hence, it is safe to clip the predictions, as well as the bounds at zero. By setting negative values to 0 the predictions and bounds are getting closer to the true k-distances, which means that the bounds get tighter, and therefore better candidate set sizes might be accomplished.

\subsubsection{Monotonicity}
The k-distances for a specific data point are monotonously increasing in $k$. Hence, we can exploit this property to derive additional bounds: 
Given a guaranteed upper bound $ub(p,k)$ with $p \in D$ and $k \in \{1, \ldots, k_{max}\}$ one can derive further guaranteed upper bounds from $k' \geq k$ as
$
    nndist(p,k) \leq ub(p,k')
$.
We can combine all bounds to obtain a tighter bound:
$
ub^*(p,k) := \min_{\substack{k' \geq k}} (ub(p,k'))
$.
This consideration analogously is valid for the lower bound. 
Since the monotonicity restoration requires bounds for different values of $k'$ we need to perform additional computations here.
However, many Machine Learning models support batched processing which allows for faster evaluation than doing the predictions sequentially.

\subsubsection{Sample Weights}
Depending on the distribution of distances for the current point to all other points, an error in the prediction, and the resulting looser bounds, have a differently severe impact on the candidate set size (cf. the introductory example in Figure~\ref{fig:skd-ol-na-256}).
Intuitively, if the point lies in a dense vicinity, increasing the upper bound might induce a significantly increased candidate set size compared to a point in a peripheral region with few points in close distance.
Thus, we propose to use sample weights to steer the model's optimization to focus more on such dense regions, and less on remote points.
Based on the observation of the effect of density, we may use a density measure to obtain sample weights.

\begin{algorithm}
\begin{algorithmic}
\State $w[i,k] = 1$ for all $i=1,\ldots,|D|$, $k=1,\ldots,K$
\For{$i=1,\ldots,ITER$}
\State Optimize model parameters $\theta$ to minimize
\State\qquad$L = w[i,k] \cdot \mathcal{L}(M(x_i, k; \theta), nndist(x_i, k))$
\State Calculate candidate set size CSS for all $x_i \in D$, $k=1,\ldots,K$
\State Update weights $w[i, k] \gets CSS(x_i, k)$
\EndFor
\end{algorithmic}
\caption{Training procedure with iterative sample re-weighting.}
\label{alg:train}
\end{algorithm}
However, since ultimately we are interested in the candidate set size, we can also directly use the candidate set size as sample weight.
To this end, we propose an iterative training scheme, given in Algorithm~\ref{alg:train}.
Starting with uniform weights, we train the model, and calculate the candidate set sizes.
Next, we use these candidate set size as sample weight, and re-train the model.
Thus, the training procedure focuses stronger on the loss from those data points where the candidate set size was large.
Here, we propose to repeat this procedure for a fixed number of steps.
However, we may also run it until some convergence criterion on the CSS is reached.
The adaptive re-weighting scheme shares some similarity with e.g. AdaBoost \cite{DBLP:conf/eurocolt/FreundS95}, but does not directly use the prediction error, but instead a different measure of interest (candidate set size) as weight.
Also the update is not multiplicative.
We also notice that the sample weights may also be used to focus on certain data points or values of $k$ stronger than on others, for instance because these values are queried more frequently.
We leave an exploration of such scenarios as future work.
 
\section{Experiments}
\label{sec:experiments}
\subsection{Experimental Setup}
The experiments are implemented in a unified pipeline to ensure a consistent evaluation basis. 
For our extensive experiments we use the Python library $\emph{hyperopt}$\cite{Bergstra_2015} for hyperparameter and model search using a random search strategy.
We implement the models using the $\emph{sklearn}$\cite{scikit-learn} and $\emph{PyTorch}$\footnote{\url{https://pytorch.org/}} libraries.  

For data preprocessing, we apply dimension-wise z-score normalization to the input, i.e. we subtract the mean and divide by the standard deviation.
We account for these additional parameters in the model size ($\mathcal{O}(d)$).
We notice that for some models we might be able to merge them with the model weights, e.g. for a neural network, we can merge the weights of the first linear layer with the linear transformation of the z-score normalization to save some parameters and reduce the index size.
Moreover, for each $k$ we normalize the $k$-distances of all points between 0 and 1.
Again, we account for these additional normalization parameters in the model size ($\mathcal{O}(k_{max})$).

\subsection{Evaluation}
For query processing two metrics are of primary importance: \emph{Runtime} and \emph{Memory consumption}. Within filter-refinement frameworks, the runtime is often approximated using the \emph{candidate set size} passed on to the refinement step.
Moreover, we follow the argumentation of \cite{10.1007/s10115-016-1004-2} that directly comparing times requires careful measurements and is of limited value, since it compares implementations\footnote{of the algorithm itself, but also the used libraries, operating system, \ldots} instead of algorithms.
The memory consumption of a model is approximated by the \emph{number of its parameters} since it is platform and programming language independent.

\begin{table}\centering
\caption{
Datasets used for evaluation.
}
\begin{tabular}{@{}llrr@{}}
\toprule
Dataset & Short name & Dimension & Size \\ 
\midrule
Oldenburg & OL & 2 & 6.105 \\
California & CAL & 2 & 21.049 \\
North America & NA & 2 & 175.814 \\
\midrule
FastText EN & EN & 300 & 200.000 \\
\bottomrule
\end{tabular}
\label{tab:datasets}
\end{table}
We evaluate on three publicly available road network datasets\cite{Li2005OnTP}\footnote{\url{https://www.cs.utah.edu/~lifeifei/SpatialDataset.htm}}, as well as a high-dimensional dataset of word embeddings of the English language \cite{DBLP:conf/iclr/LampleCRDJ18}\footnote{\url{https://github.com/facebookresearch/MUSE}}.
Their statistics are summarized in Table~\ref{tab:datasets}. 
We compare with MRkNNCoP tree \cite{Achtert:2006:ERK:1142473.1142531} (CoP) since it is a state-of-the-art approach based on the filter-refinement framework.
Moreover, it relies on a linear approximation in log-log space and hence serves as ablation to show the effect of allowing arbitrary-shaped approximations.
For the experiments, the following model types are used: Decision Trees, GradientBoosting Regressors, AdaBoostRegressors and Neural Networks, with the hyperparameters optimized in the following ranges:

\begin{itemize}
    \item Decision Tree: $max\_depth \in \{1, 2, \ldots, 15\}$,
    \item Gradient Boosting: $n\_trees \in \{1, 2, \ldots, 500\}$, $max\_depth \in \{1, 2, \ldots, 15\}$, $\eta \in [0.0, 1.0]$,
    \item AdaBoost: $n\_trees \in \{1, 2, \ldots, 500\}$, $\eta  \in [0.0, 1.0]$,
    \item Neural Network: $n\_layers \in \{1, \ldots, 5\}$, $units(i) \in \{4, \ldots, 300\}$, $batch\_norm \in \{True, False\}$, $batch\_size  \in \{2^6, \ldots, 2^{12}\}$, $dropout\_rate \in [0.0, 1.0]$, $input\_norm \in \{True, False\}$, $loss \in \{MAE, MSE\}$
\end{itemize}
where $\eta$ denotes the learning rate.
We always used four iterations of sample re-weighting.

\subsection{Trade-off between Candidate Set Size and Model Size}
\label{subsec:cs-eval}

\begin{figure}
\centering
\includegraphics[width=.7\linewidth]{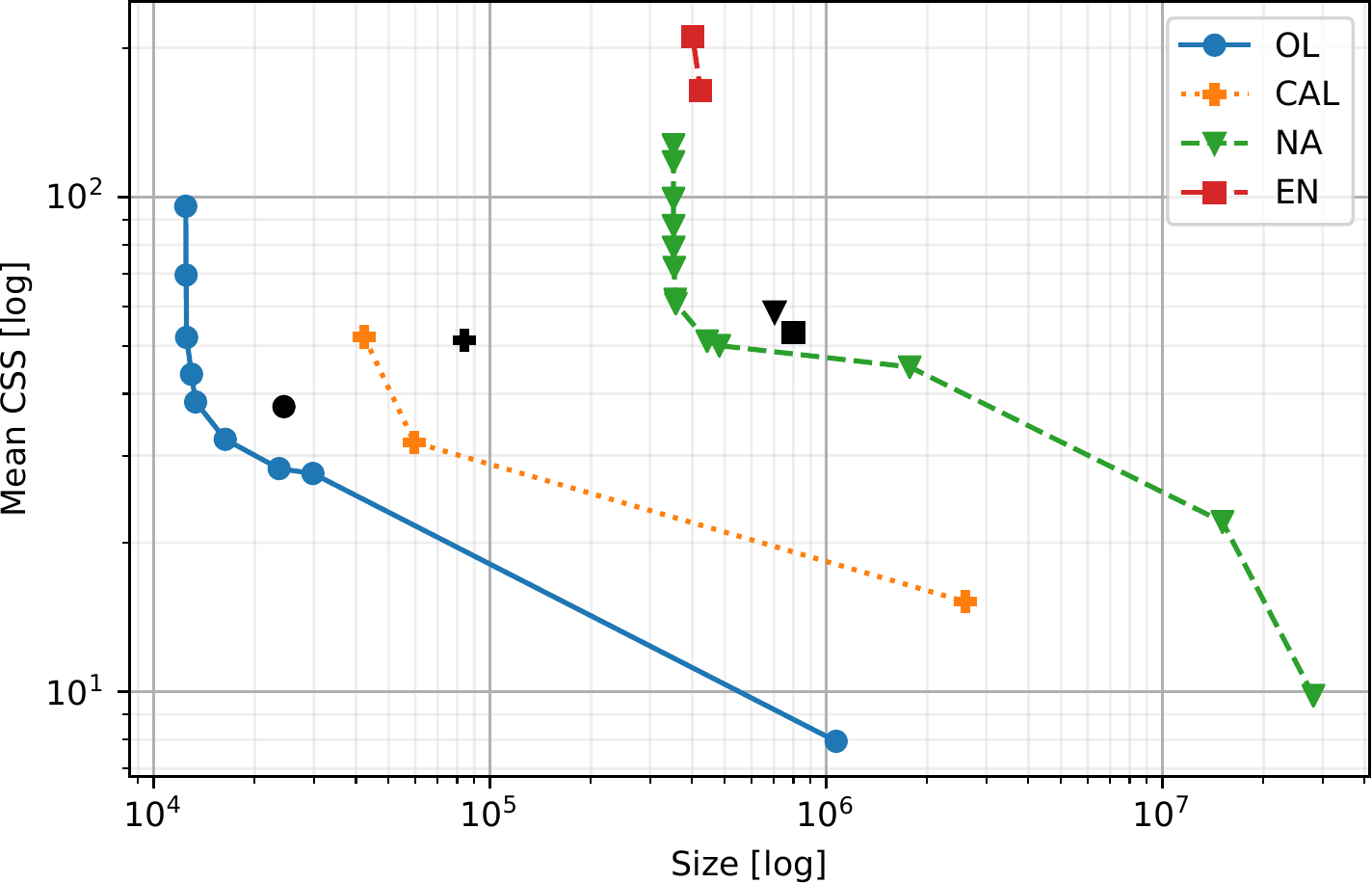}
\caption{
Trade-off between mean candidate set size (CSS) and model size (both on logarithmic scale).
For each dataset, we only show the pareto-optimal models. In black, and the corresponding symbol, we show the performance of MRkNNCoP tree (CoP).
For most datasets, our approach outperforms CoP in both, model size and CSS.
For the EN dataset, we obtain smaller models, but exceed the CSS of CoP.
}
\label{fig:overall-performance-mean-cs}
\end{figure}

\begin{figure}
\centering
\includegraphics[width=.7\linewidth]{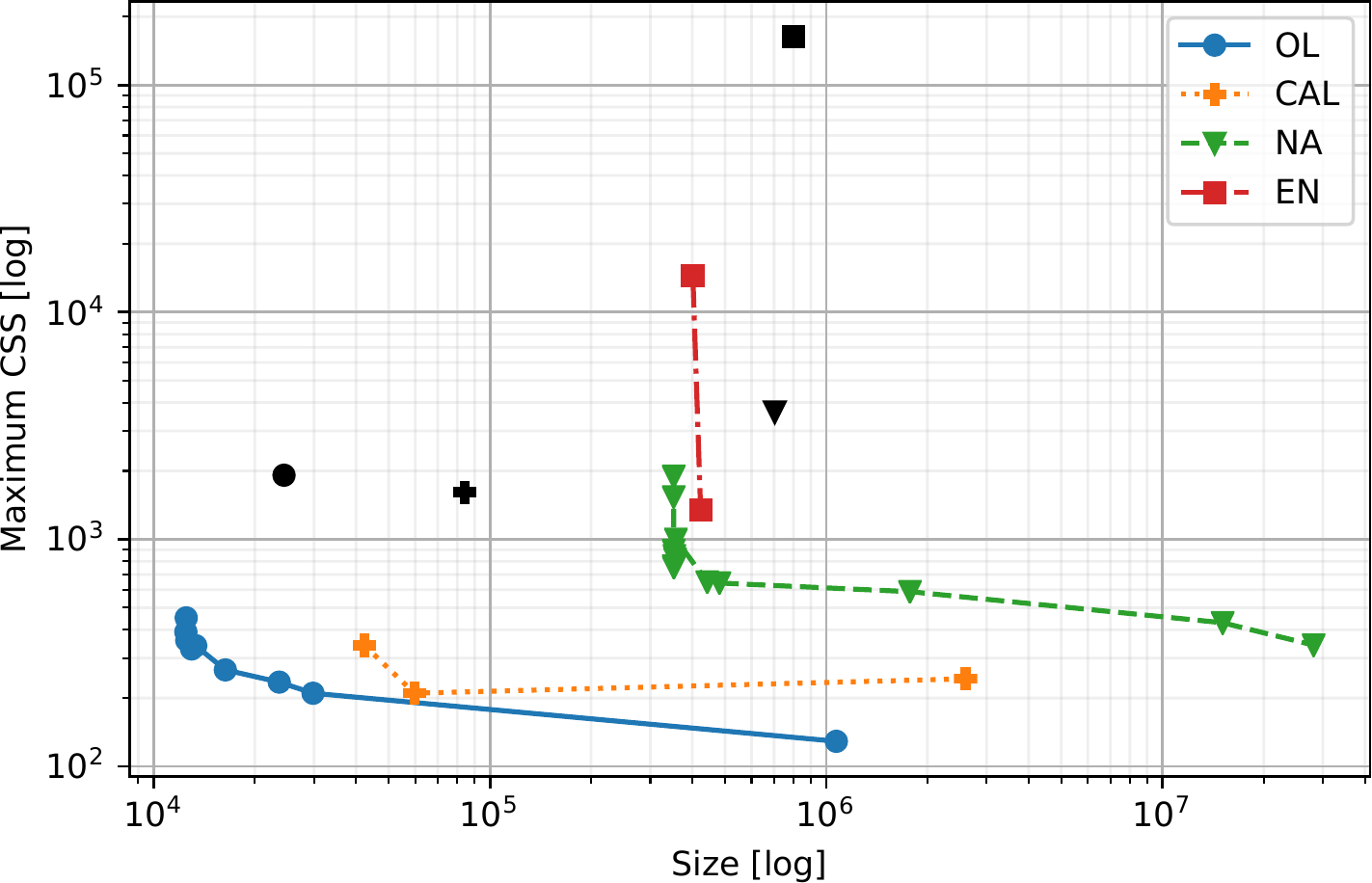}
\caption{
Maximum candidate set size (CSS) compared to the model size for the same models as in Figure~\ref{fig:overall-performance-mean-cs}, again both on a logarithmic scale.
Different colors/markers indicate different datasets, the black symbols show the performance of MRkNNCoP tree (CoP).
For all datasets we can strongly improve the maximum CSS compared to CoP, reducing the worst-case runtimes and thus mitigating latency spikes in query processing.
\label{fig:overall-performance-max-cs}
}
\end{figure}
For brevity, we abbreviate candidate set size with CSS for the remainder of this paper.
Figure~\ref{fig:overall-performance-mean-cs} shows the overall performance of our models for each dataset compared to CoP.
Each marker type and color indicates a different dataset.
The lines indicate the skyline of our models regarding their achieved mean CSS and sizes, i.e. we only show the Pareto-optimal models, where there does not exist a model which is smaller \emph{and} has a smaller mean CSS at the same time. 
The black single markers stand for the results of MRkNNCoP for each dataset. 
Both axes use a logarithmic scale.
For all spatial datasets (OL, CAL, NA) our models outperform CoP in both, the mean CSS as well as the model size. 
For the word embeddings (EN) the mean CSS of CoP is smaller, however, we can reduce upon the model size.

In Figure~\ref{fig:overall-performance-max-cs}, we additionally show the same models, but now comparing \emph{maximum} CSS against model size.
Here, our models improve over CoP on all datasets regarding both metrics, up to several orders of magnitude.
The maximum CSS correlates with the worst-case runtime of a query, which in practice may cause latency spikes in query processing.

From the studied model types only decision trees and neural networks occur in the skyline of any dataset.
EN's skyline only contains neural networks.
For CAL and OL, the largest model in the skyline is a decision tree, and the rest comprises only neural networks.
For NA, the two largest models are decision trees.

\subsection{Ablation Study}
\begin{table*}
\centering
\caption{
Ablation study showing the mean candidate set size ($\overline{CSS}$), maximum candidate set size ($\widehat{CSS}$) and the model size.
With \emph{S} we denote whether we use sample weights, with \emph{K} and \emph{D} whether we aggregate over all $k$ values, or over all points (or both), and by \emph{M} we denote whether we restore monotonicity. 
As base we choose the model with the best $\overline{CSS}$ which is still smaller than the MRkNNCoP tree, which is usually situated in the middle of the skyline.
}
\label{tab:ablation}
{\scriptsize
\begin{tabular}{cccc|rrr|rrr|rrr|rrr}
\toprule
\multicolumn{4}{c}{dataset} & \multicolumn{3}{|c|}{OL} & \multicolumn{3}{c}{CAL} & \multicolumn{3}{|c|}{NA} & \multicolumn{3}{c}{EN} \\
S & K & D & M & $\overline{CSS}$ & $\widehat{CSS}$ &   Size & $\overline{CSS}$  & $\widehat{CSS}$  &   Size & $\overline{CSS}$ & $\widehat{CSS}$  &   Size & $\overline{CSS}$  & $\widehat{CSS}$  &   Size \\
\midrule
 \checkmark &  \checkmark &  \checkmark &  \checkmark & \textbf{26.29} &           118 &           24,147 & \textbf{31.46} & \textbf{168} &           60,185 & \textbf{50.04} &           642 &           482,052 &           163.61 &           1,457 &          423,749 \\
  \checkmark &  \checkmark &  \checkmark &             &           26.39 &           119 &           24,147 &           31.55 & \textbf{168} &           60,185 &           50.06 &           642 &           482,052 &           163.66 &           1,457 &          423,749 \\
  \checkmark &  \checkmark &             &  \checkmark &           28.32 &           209 &           23,635 &           31.89 &           210 &           59,673 &           50.05 &           642 &           481,540 &           164.37 &           1,457 &          423,237 \\
  \checkmark &  \checkmark &             &             &           28.34 &           209 &           23,635 &           31.95 &           210 &           59,673 &           50.06 &           642 &           481,540 &           164.42 &           1,457 &          423,237 \\
  \checkmark &             &  \checkmark &  \checkmark &           49.65 &           143 & \textbf{11,937} &           66.53 &           243 & \textbf{18,089} &          423.14 &         1,858 & \textbf{130,426} &          5334.82 &          36,625 & \textbf{23,749} \\
  \checkmark &             &  \checkmark &             &           52.37 &           164 & \textbf{11,937} &           68.25 &           262 & \textbf{18,089} &          428.59 &         1,939 & \textbf{130,426} &          5351.05 &          36,625 & \textbf{23,749} \\
             &  \checkmark &  \checkmark &  \checkmark &           27.07 & \textbf{117} &           24,147 &           31.80 &           182 &           60,185 &           51.32 & \textbf{594} &           482,052 & \textbf{163.34} & \textbf{1,350} &          423,749 \\
            &  \checkmark &  \checkmark &             &           27.18 &           118 &           24,147 &           31.88 &           186 &           60,185 &           51.34 & \textbf{594} &           482,052 &           163.40 & \textbf{1,350} &          423,749 \\
            &  \checkmark &             &  \checkmark &           29.16 &           194 &           23,635 &           32.36 &           221 &           59,673 &           51.32 & \textbf{594} &           481,540 &           164.03 & \textbf{1,350} &          423,237 \\\
            &  \checkmark &             &             &           29.18 &           194 &           23,635 &           32.41 &           221 &           59,673 &           51.34 & \textbf{594} &           481,540 &           164.08 & \textbf{1,350} &          423,237 \\
            &             &  \checkmark &  \checkmark &           52.56 &           151 & \textbf{11,937} &           68.85 &           250 & \textbf{18,089} &          480.64 &         2,460 & \textbf{130,426} &          5201.22 &          37,236 & \textbf{23,749} \\
            &             &  \checkmark &             &           55.93 &           183 & \textbf{11,937} &           70.11 &           251 & \textbf{18,089} &          497.34 &         2,460 & \textbf{130,426} &          5215.65 &          37,236 & \textbf{23,749} \\
\bottomrule

\end{tabular}
}
\end{table*}

To investigate the effect of our models' components, we perform an ablation study.
For each dataset we chose the best model according to mean candidate set size (CSS) which is still smaller than the MRkNNCoP tree as the base model.
For this model, we study the effect of disabling sample weights, aggregating only over $k$, or the points, and turning off the monotonicity enhancement.
We record mean and maximum CSS, as well as model size.
The overall results are summarized in Table~\ref{tab:ablation} and discussed in the following.

\subsubsection{Effect of Sample Weights}
When comparing the usage of sample weights, we observe that the mean CSS always improves for all road network datasets (OL, CAL, NA) irrespective of the base configuration, e.g. whether monotonicity restoration is used, etc.
We observe the strongest absolute improvement for NA, ($K \to SK$, $497.34 \to 480.64$).
For the word embedding dataset (EN), we observe the opposite effect:
Here, it is advantageous to not use sample re-weighting, although the difference is usually small.

\subsubsection{Effect of Residual Aggregation Axes}
Across all datasets, we observe that aggregation over $k$ (i.e. one fixed bound width per data point $p$) is advantageous compared to the aggregation of $p$ with one fixed bound width per value of $k$.
This is likely because the points reside in differently dense areas and sharing the same bound width over all data points does not take this into account.
Since $k_{max} \ll n$, the bounds obtained by aggregation over $k$  require more memory ($\mathcal{O}(n)$) than those by aggregation over $p$ ($\mathcal{O}(k)$).
By combining both bounds we obtain a guaranteed improvement in candidate set size with slightly increased index size $\mathcal{O}(n+k)$.
Thus this parameter allows us to steer the trade-off between index size and candidate set size.
When already spending $n$ parameters for bounds aggregated over $k$, we suggest to also combine them with bounds aggregated over $p$ since for usual values of $k_{max} \ll n$ they offer a guaranteed (but potentially minor) improvement at negligibly increased index size.

\subsubsection{Effect of Monotonicity Restoration}
\begin{figure}
\centering
\includegraphics[width=.7\linewidth]{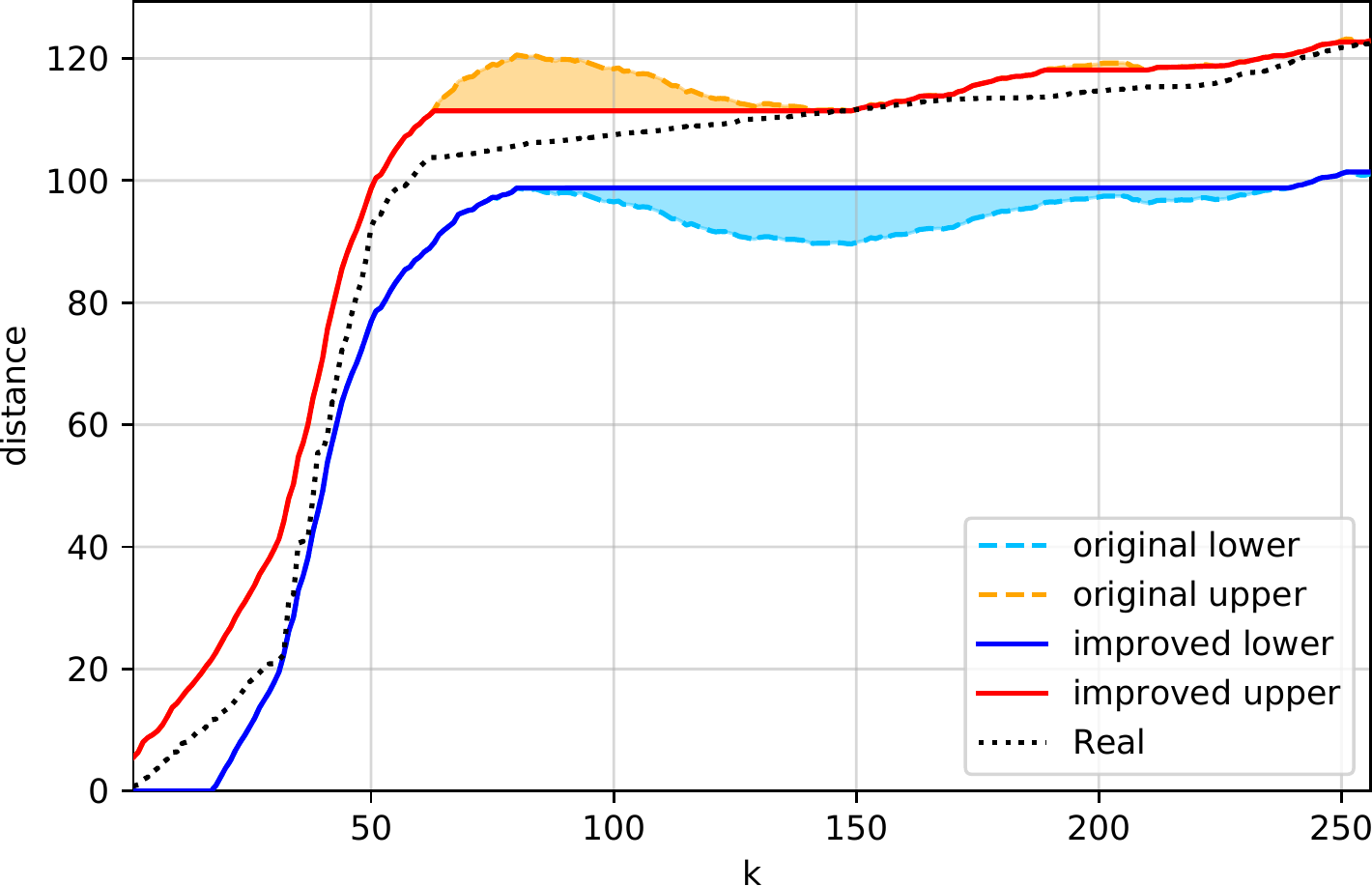}
\caption{
Qualitative example of restoring monotonicity for upper and lower bound on the dataset NA. 
The actual model bounds (dashed, lower: blue, upper: red), here obtained by aggregation over $k$, and the improved bounds (solid, lower: blue, upper: red) are displayed for each $k$ (x-axis). 
The areas filled in the corresponding color could be eliminated by restoring monotonicity.
}
\label{fig:effect-mono-k}
\end{figure}

If the residuals are aggregated only over $p$, the original bounds \emph{before} bound enhancement have a fixed width only dependent on $k$.
If the residuals are aggregated only over $k$, then the original bounds \emph{before} bound enhancement even have a fixed width over all $n$.
However, there may be differences \emph{after} bound enhancement, as it depends on the distribution of the predictions for a specific data point $p$ concerning $k$, which is different for each data object. 
From Table~\ref{tab:ablation} we can observe that monotonicity restoration has the largest effect when applied to bounds obtained only from aggregation over $p$.
To better illustrate this effect we show a qualitative example from the NA dataset in Figure~\ref{fig:effect-mono-k}, where restoring monotonicity improves the tightness of the bounds.
The real $k$-distances are shown as a black dotted line.
With dashed lines we show the original upper (red) and lower (blue) bounds predicted by the model. 
The solid lines indicate the monotonicity bounds after bound enhancement. 
The colored areas between the original bound and the enhanced bound could be eliminated by restoring monotonicity. 
As the bound width is fixed over all $k$ (since we aggregated over this axis), as soon as the model prediction is non-monotonous, both bounds become non-monotonous as well. 
\section{Conclusion}
\label{sec:conclusion}
In this paper we proposed a novel approach to approximate the $k$-distances directly using a Machine Learning model in order to obtain filters for RkNN query processing within the filter-refinement framework.
To this end, we formalized the prediction as a regression problem and derived several guaranteed bounds essential to guarantee the completeness of the filter step.
Moreover, we showed how to further enhance the bounds by restoring non-negativity and monotonicity.
We also proposed an iterative method to obtain sample weights to further tweak our models' candidate set sizes.
In experiments on several real-world datasets, we showed superior performance to a state-of-the-art method, the MRkNNCoP tree, not only in the number of candidates but also regarding the index size.

\bibliographystyle{IEEEtran}
\bibliography{main.bib}

\end{document}